\documentclass[article,moreauthors,pdftex,10pt,a4paper]{mdpi}
\usepackage{upgreek}
\usepackage{subcaption}

\Title{Towards an Improved Test of the Standard Model's Most Precise Prediction}

\Author{G. Gabrielse $^{1*}$, S. E. Fayer $^{1}$, T. G. Myers $^{1}$ and X. Fan $^{1,2}$}

\AuthorNames{G. Gabrielse, S. E. Fayer, T. G. Myers, and X. Fan}

\address{%
$^{1}$ \quad Center for Fundamental Physics, Northwestern University\\
$^{2}$ \quad Dept. of Physics, Harvard University\\
}

\corres{Correspondence: gerald.gabrielse@northwestern.edu;}

\abstract{The electron and positron magnetic moments are the most precise prediction of the standard model of particle physics.  The most accurate measurement of a property of an elementary particle has been made to test this result. A new experimental method is now being employed in an attempt to improve the measurement accuracy by an order of magnitude. Positrons from a "student source" now suffice for the experiment.  Progress toward a new measurement is summarized.}


\begin{document}


\newcommand{\mue}{\mu_{e-}}
\newcommand{\mupos}{\mu_{e+}}
\newcommand{\muboth}{\mu_{e\pm}}
\newcommand{\mueVector}{\vec{\boldsymbol{\mu}}_{e-}}
\newcommand{\muposVector}{\vec{\boldsymbol{\mu}}_{e+}}
\newcommand{\mubothVector}{\vec{\boldsymbol{\mu}}_{e\pm}}
\newcommand{\gpos}{g_{e+}}
\newcommand{\gboth}{g_{e\pm}}

\section{Introduction}

The Standard Model of Particle Physics (SM) is the great triumph and the great frustration of modern physics \cite{StandardModelTriumph}.  It is a great triumph because all laboratory tests of the SM have so far agreed with its predictions within measurement and calculation uncertainties.  It is a great frustration because the SM and its symmetries are unable to account for basic properties of our universe.  According to the SM, for example, the big bang would produce essentially equal numbers of matter and antimatter particles which would then annihilate as the universe cools.  Yet, we do have a universe and it is composed of matter rather than antimatter for reasons that are not understood.  The SM also does not fit gravity very well and cannot account for inflation, dark matter or dark energy.  

The most precise prediction of the SM is the size of the electron magnetic moment  $\mueVector$.  The intrinsic CPT symmetry of the SM predicts that the positron magnetic moment $\vec{\boldsymbol{\mu}}_{e+}$ should be opposite in sign and equal in magnitude.  At the current level of measurement precision, the Dirac contribution, quantum electrodynamics through the tenth order, and hadronic contributions are all critical.  At the proposed new measurement accuracy, weak interactions effects will also be important.  All these contributions from the SM are tested by comparing the prediction to measurements.  This is already the most precise confrontation of theory and experiment and this seems likely to continue to be true.

The spin 1/2 electron and positron are eigenstates of spin $\vec{\mathbf{S}}$ with eigenvalue $\hbar/2$.  Their magnetic moments  can be written as 
\begin{equation}
   \mubothVector = \muboth \, \frac{\vec{\mathbf{S}}}{\hbar/2},
    \label{eq:gDef}
\end{equation}
where the upper and lower signs designate the positron and electron, respectively.   Comparing $\muposVector$ to $\mueVector$ is a test of the standard model's CPT invariance which, as mentioned, predicts that they should have equal magnitudes and opposite signs.

The natural size scale of the electron and positron moments is the Bohr magneton, $\mu_B = e\hbar/(2m)$, where $-e$ is the electron's charge, and $m$ is its mass. An electron in a circular orbit with angular momentum $\hbar$ has a magnetic moment $-\mu_B$.  The positron moment for the same orbit is opposite in sign because the positron orbits in the opposite direction.  The electron and positron moments in Bohr magnetons are sometimes written in terms of their g-values, defined to be the positive values 
$\gboth/2 = \pm \muboth/\mu_B$.

An evaluation of the SM prediction for the electron and positron moments is in Eq.~\ref{eq:StandardModelPrediction}.The calculations and computations needed to make this prediction are summarized in a following section and in much more detail in \cite{gfactortheoryNio2019}. The measured $\alpha$ with the lowest uncertainty \cite{AlphaCs2018} is an input for the prediction.
\begin{align}
SM~Prediction~with~Cs~2018:&     &     -\mue/\mu_B = \mupos/\mu_B = \, 1.001 \, 159 \, 652 \, 181 \, 61 \, (024) \qquad[0.24~\rm{ppt}],&
\label{eq:StandardModelPrediction}\\
Measured:&     &    -\mue/\mu_B = \, 1.001 \, 159 \, 652 \, 180 \, 73 \, (028) \qquad [0.28~\rm{ppt}],&
\label{eq:ElectronMagneticMoment}\\
Measured:&   &   \quad \mupos/\mu_B = \, 1.001 \, 159 \, 652 \, 187 \, 90 \, (430) \qquad[4.30~\rm{ppt}].&
\label{eq:PositronMagnetic Moment}
\end{align}
(In the latter, the zero to the right in the value and in the uncertainty are added to allow easy comparison of  Eqs.~\ref{eq:StandardModelPrediction}-\ref{eq:PositronMagnetic Moment}).  The most accurate measurement of the electron magnetic moment, made in our laboratory \cite{HarvardMagneticMoment2008}, is in Eq.~\ref{eq:ElectronMagneticMoment}. It is a quantum measurement made by observing quantum jumps between the quantum energy levels of one electron suspended in a Penning trap.  The most accurate positron measurement, made at the University of Washington \cite{DehmeltMagneticMoment}, was not a fully quantum measurement and it has a 15 times larger uncertainty. In the next section we discuss the remarkable agreement of the prediction and measurement, and the possibility that there may be a small but significant disagreement.    

The goal of our new measurement is to reduce the uncertainty in the measured electron moment by a factor of 10 -- an important step towards testing the standard model's most precise prediction an order of magnitude more stringently.  Because the positron moment has been measured less precisely, the positron measurement that we propose to perform at the same accuracy would improve the precision at which the positron and electron are compared, to test the standard model's CPT invariance, by at least a factor of 150.  Until a new measurement of the fine structure constant is carried out, the more accurate electron magnetic moment will determine the fine structure constant 10 times more accurately than it is currently known.  

As long as the standard model remains the great triumph and the great frustration of modern physics, it is very important to test just how accurate is the standard model's most precise prediction, and to test just how exact is the standard model's fundamental symmetry.

The following sections summarize the compelling motivations for a new measurement of the electron and positron magnetic moments. 
\begin{itemize}
    \item Testing the standard model's most precise prediction (Section \ref{sec:TestingSM})
    \item Most stringent test of CPT invariance with leptons (Section \ref{sec:TestingCPT})
    \item Most accurate determination of the fine structure constant (Section \ref{sec:DeterminingAlpha})
    \item Probe for electron substructure (Section \ref{sec:TestingSubstructure})
    \item Comparison to the muon magnetic moment (Section \ref{sec:Muon})
    \item Additional implications of the magnetic moment measurement (Section \ref{sec:Other})
\end{itemize}
A complete description of the apparatus and methods being developed to allow a more accurate measurement of the electron and positron magnetic moments would require a much longer discussion.  The key elements discussed here include:
\begin{itemize}
    \item A new quantum measurement (Section \ref{sec:BasicIdea})
    \item The key new method of a simultaneous determination of the cyclotron and spin frequencies (Section \ref{sec:Simultaneous})
    \item A new trap and refrigerator apparatus (Section \ref{sec:Apparatus})
    \item Gas $^3$He NMR probe developed to improve the spatial stability and measure magnetic field stability (Section \ref{sec:NMR})
    \item Microwave trap cavity for cavity sideband cooling (Section \ref{sec:TrapCavity})
    \item Safe and efficient positron loading (Section \ref{sec:PositronLoading})
\end{itemize}
Hopefully the considerable effort going into developing the new apparatus and methods will result in measuring the electron magnetic moment 10 times more accurately and in comparing the electron and positron 150 times more accurately.

\section{Testing the Standard Model's Most Precise Prediction}
\label{sec:TestingSM}

The standard model's most precise prediction is  that the electron magnetic moment is given by  
\begin{align}
	-\frac{\mu_-}{\mu_B} = \frac{\mu_+}{\mu_B}  =1 &+ C_2\left(\frac{\alpha}{\pi}\right) +C_4\left(\frac{\alpha}{\pi}\right)^2 +C_6\left(\frac{\alpha}{\pi}\right)^3
	+C_8\left(\frac{\alpha}{\pi}\right)^4 +... 	 + a_{\text{hadronic}}+a_{\text{weak}},  \label{eq:StandardModelSeries}
\end{align}
while the positron moment is the same with opposite sign.
The leading contribution is the $1$ that is predicted for a Dirac point particle.  Vacuum fluctuations and polarization modify the interaction of the
electron with the magnetic field, increasing the effective magnetic moment of the electron by approximately one part per thousand.  This addition is
described by the infinite QED series in powers of $\alpha/\pi$, where $\alpha$ is the fine structure constant approximately equal to $1/137$.  The two remaining terms are the hadronic and weak interaction contributions.

Substantial $n$-vertex QED calculations are needed to determine the coefficients, $C_n$.   The current state of the exact and numerical evaluations of these are summarized in \cite{gfactortheoryNio2019}.  We summarize what is needed to compare measurements and predictions.  The first four coefficients ($C_2$, $C_4$, $C_6$ and $C_8$) have all been calculated exactly except that they depend weakly upon ratios of lepton masses, for which we use the values in Particle Data Group compilation of 2018 \cite{PDG2014paper}. Laporta's very recent analytic calculation of 891 Feynman diagrams produced an exact $C_8$  \cite{LaportaC8} that supersedes and agrees with a substantial numerical calculation \cite{Nio2018}.  An impressive numerical calculation by Nio and Kinoshita of the $12672$ Feynman diagrams needed for $C_{10}$ continues to get more precise as better Monte Carlo integration statistics accumulate \cite{gfactortheoryNio2019}.  The current values of these coefficients are
\begin{align}
    C_2 =& \, \quad 0.5\\
    C_4 =& -0.328 \, 478 \, 444 \, 002 \, 62 \, (25)\\
    C_6 =& \, \quad 1.181 \, 234 \, 016 \, 818 \, 3 \, (79)\\
    C_8 =& -1.911 \, 321 \, 391 \, 8 \, (12) \\
    C_{10} =& \, \quad 6.73 \, (16)
\end{align}
with the uncertainty in the last digits in parentheses.   The hadronic and weak contributions
\begin{align}
    a_{hadronic} =& \quad 1.693 \, (11) \times 10^{-12}\\
    a_{weak} =&  \quad 0.030 53 \, (23) \times 10^{-12}
\end{align}
have been estimated with their uncertainties \cite{aHadronicWeak2017}.

\begin{figure}[htbp!]
	\centering
	\includegraphics*[width=4.8in]{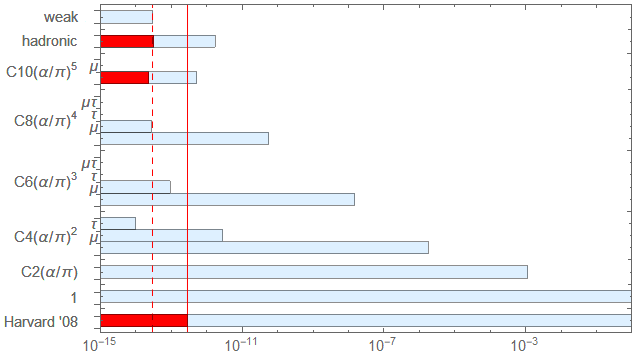}
	\caption{Contributions to the calculated SM value for $\pm \muboth/\mu_B$ with uncertainties as red bars. The solid and dashed red lines are the uncertainty for our completed and proposed measurements \cite{HarvardMagneticMoment2008}. }  \label{fig:SMuncertain}
\end{figure}

To illustrate the standard model contributions that are probed by our completed and proposed measurements of the electron $\mue/\mu_B$, Fig.~\ref{fig:SMuncertain} gives a graphical comparison of the standard model contributions and the uncertainty in the completed (solid line) and proposed (dashed line) measurements. 
The lower bar represents the total measured electron moment, with the red portion (and the red line) indicating the uncertainty in our 2008 measurement.  The dashed line indicates the ten times improved uncertainty to which we aspire.  

The Dirac contribution to the moment is the second bar, which has no uncertainty.  The quantum electrodynamics contributions are the next 5 bars.  There is no computational uncertainty in these contributions because they have been calculated analytically.  The only uncertainty in these contributions, too small to show in the figure, come because those marked $\mu$ and $\tau$ depend upon the measured ratios of these leptons and the electron masses. The 10th order contribution has a numerical uncertainty from the Monte Carlo integrations. The 12th order contribution is too small to contribute unless the completely unknown $C_{12}$ turns out to be exceedingly large, since $\alpha/\pi \approx 0.002$ and $(\alpha/\pi)^6$ would not be visible in the figure.  The hadronic contribution 
is already very important at the current experimental uncertainty, and even more so at the proposed precision.  Its estimated uncertainty is low enough to allow testing the standard model prediction ten times more precisely.  The weak force contribution is below the current experimental precision and will just be important at the proposed uncertainty in the new measurement.

Evaluating the standard model prediction in Eq.~\ref{eq:StandardModelPrediction} requires a measured value of the fine structure constant, 
\begin{equation}
\alpha \equiv \frac{e^2}{4\pi\epsilon_0 \hbar c} \approx \frac{1}{137},
\label{eq:alphaintro}
\end{equation}
that is the strength of the electromagnetic coupling in the low-energy limit. For some time, $\epsilon_0$ has been a defined constant, while $e$ and $\hbar$ are measured quantities that cannot be calculated.  Neither was measured accurately enough to determine $\alpha$ to the precision of the methods we will discuss. As of 20 May 2019, $e$ and $\hbar$ will be redefined to be exact \cite{SIredef2019}.  Measuring $\alpha$ will thus become equivalent to measuring $\epsilon_0$, since the latter will no longer be exactly defined.  

For testing the standard model,  $\alpha$ must be determined independently of Eq.~\ref{eq:StandardModelPrediction}. The most accurate independent determinations now use Rb or Cs atoms with the alternative expression,
\begin{equation}
    \alpha^2 = \frac{2 R_\infty}{c} \frac{A(x)}{A(e)} \frac{h}{M(x)}.
    \label{eq:AlternateAlpha}
\end{equation}
Separate measurements are made of the Rydberg constant ($R_\infty$), the atomic mass of the electron ($A(e)$), the atomic mass of either a Rb or Cs atom ($A(x)$), and $h/M(x)$ for the same atom.  The speed of light, $c$, is a defined value. Because of the square in Eq.~\ref{eq:AlternateAlpha}, half of the fractional uncertainty in each of the measured quantities is the contribution to the uncertainty in $\alpha$. 

Hydrogen spectroscopy of the 1s-2s transition \cite{Haensch1s2s2011} and a higher transition \cite{Beyer2017Rydberg,RydbergOrsay2018}, with a calculation of the hydrogen structure (that includes important QED corrections) can relate measured transition frequencies to the Rydberg constant. This gives two new 2018 values \cite{Beyer2017Rydberg,RydbergOrsay2018}: 
\begin{align}
R_\infty(MPQ) &= \, 10 \, 973 \, 731.568 \, 076 \, (096) ~m^{-1}\\
R_\infty(Orsay) &= \, 10 \, 973 \, 731.568 \, 530 \, (140) ~m^{-1}
\end{align}
The recent surprise that the first differs from the second (and others) by nearly 4 standard deviations is not very important for determining $\alpha$ since this difference would only shift $\alpha$ by $20$ ppt, while $\alpha$ gets a $200$ ppt  uncertainty from other sources.
The electron mass in amu, $A(e)$, that we use \cite{CODATA2014paper} comes primarily from comparing calculations of the magnetic moment of an electron bound to an atom, including substantial QED corrections, with a measurement of this moment \cite{Sturm2014emass}. 
\begin{align}
A(e) =& \, 0.000 \, 548 \, 579 \, 909 \, 070 \, (16) ~~~~~[29 ~ppt]\\
A(Rb) =& \,\,\,\, 86.909 \, 180 \, 5319 \, (65) ~~~~~[75 ~ppt]\\
A(Cs) =& \, 132.905 \, 451 \, 9615 \, (86) ~~~~~[65 ~ppt]
\end{align} 
The atomic masses in amu are from the AME 2016 atomic mass evaluation \cite{AME2016}, determined from the cyclotron frequency of their ions in Penning traps \cite{Mount2011Rb} \cite{PritchardMassRatios1999}. 

Almost all of the uncertainty is from the measured $h/M$ for the Rb \cite{RbAlpha2011} and Cs atoms \cite{AlphaCs2018},  
\begin{align}
h/M(Rb) = & \, 4.591 \, 359 \, 272 \, 9 \, (57) \times  10^{-9} ~m^{2}/s ~~~~~[1200 ~ppt]\\
h/M(Cs) = & \, 3.002 \, 369 \, 472\, 1 \, (12) \times 10^{-9} ~m^{2}/s ~~~~~[400  ~ppt],
\end{align}
measured using atom interferometers.  These measurements together  give
\begin{align}
\alpha^{-1}(Rb~2011) =& \, 137.035 \, 998 \, 995 \, (85) ~~~~~[620 ppt] \label{eq:RbAlpha}\\
\alpha^{-1}(Cs~2018) =& \, 137.035 \, 999 \, 045 \, (28) ~~~~~[200 ppt] \label{eq:CsAlpha}
\end{align}
for the Rb and Cs values of the fine structure constant.  
Note that extracting both $R_\infty$ and $A(e)$ from measurements required standard model QED theory, but these calculations are clearly independent of Eq.~\ref{eq:StandardModelSeries}. 

\begin{figure}[htbp!]
    \centering
    \includegraphics[width=3in]{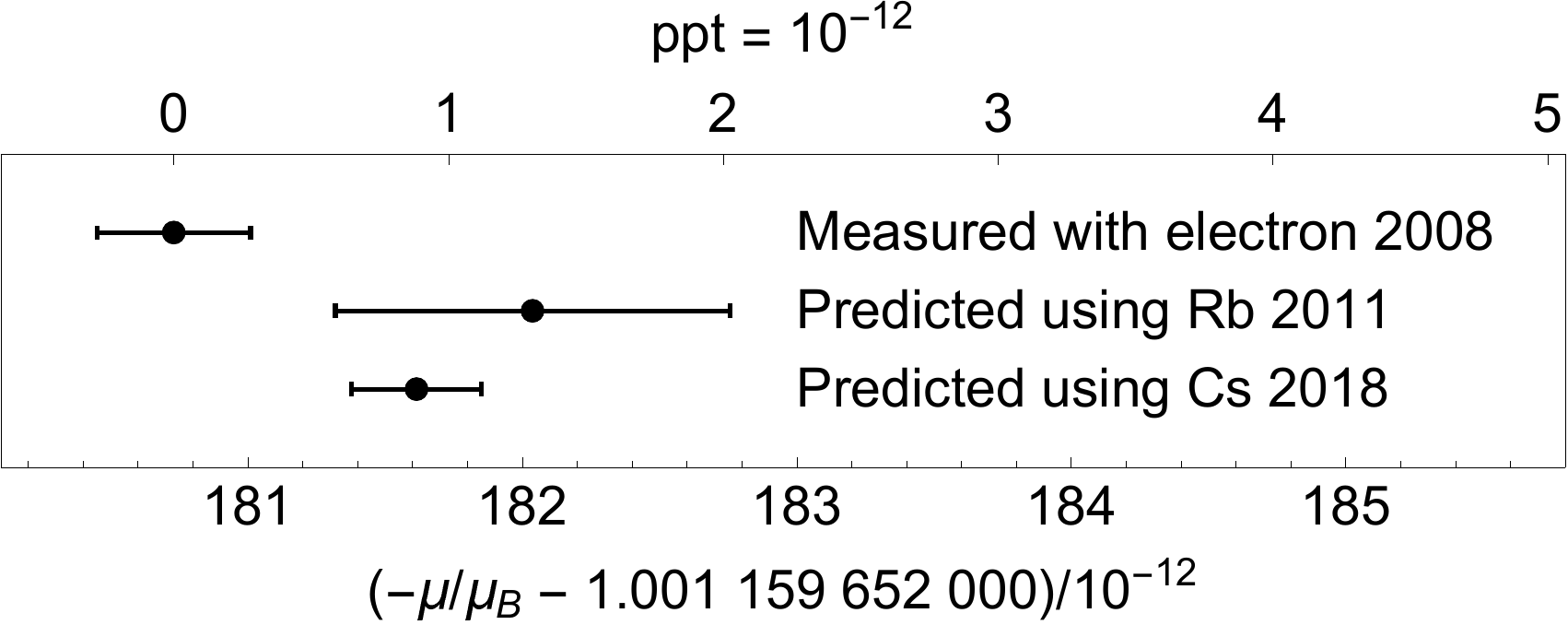}
    \caption{Comparison of the measured electron magnetic moment with the standard model predictions that use the two independent measurements of $\alpha$ that have the lowest uncertainties.}
    \label{fig:Triumph}
\end{figure}

The electron magnetic moment predicted by the standard model has an uncertainty that comes from that in the calculation of the $C_n$, $a_{hadronic}$ and $a_{weak}$, along with the uncertainty in the measured $\alpha$. Thanks to the recent theoretical progress, the prediction uncertainty is almost entirely from the uncertainty in $\alpha$. The predictions using the $\alpha$ deduced independently from measurements   in Eq.~\ref{eq:StandardModelPrediction}, 
\begin{align}
SM~and~\alpha(Rb~2011):~~~~~      &-\mue/\mu_B = \mupos/\mu_B = \, 1.001 \, 159 \, 652 \, 182 \, 04\, (72) \qquad[0.72~\rm{ppt}]\\
SM~and~\alpha(Cs~2018):~~~~~      &-\mue/\mu_B = \mupos/\mu_B = \, 1.001 \, 159 \, 652 \, 181 \, 62 \, (24) \qquad[0.24~\rm{ppt}].
\label{eq:StandardModelPredictionRb}
\end{align}
Fig.~\ref{fig:Triumph} compares our measured moment with the standard model predictions that utilize both of the two best independently measured values of $\alpha$. 

Arguably the greatest triumph of the standard model is that it is able to accurately predict the electron magnetic moment to about 1 ppt, the precision at which the measured and predicted moments agree. This is the first message conveyed by Fig.~\ref{fig:Triumph}. Given that the standard model cannot predict very basic features of the universe, as discussed earlier, this accomplishment of the standard model is extremely remarkable.

The second message from Fig.~\ref{fig:Triumph} is that the measured moment and the most precise predicted moment differ by 2.4 standard deviations. This difference adds motivation to our push towards a new measurement in an entirely new apparatus using new methods.  Whether or not this small difference holds up as better measurements of the electron magnetic moment and the fine structure constant are carried out remains to be seen. 

If we achieve the goal of this work, to increase the accuracy of the measured electron magnetic moment by a factor of ten, a comparably improved independent measure of $\alpha$ will be required to test the standard model.  The reported accuracy of the standard model calculations of the QED, hadronic and weak interaction terms is already sufficient.  We naturally hope that an improved measurement of the electron magnetic moment will stimulate an improved independent measurement of $\alpha$ as has happened before.

\section{Most Stringent Test of CPT Invariance with Leptons}
\label{sec:TestingCPT}

CPT symmetry is intrinsic to the field theory that is part of the standard model. This symmetry is often assumed rather than tested as a result.  In light of the fact that P and CP symmetry were once broadly assumed, and only later shown to not be universal, CPT symmetry should be tested in fundamental systems that can be easily understood and precisely measured.  Something being wrong or at least missing from the standard model and its symmetries, as discussed above, reinforces this conclusion.   

CPT invariance predicts that the magnetic moments of the electron and positron should be opposite in direction but with the same magnitude.  No standard model computations are required to test the CPT prediction.  
The best measured values are in Eqs.~\ref{eq:ElectronMagneticMoment} and \ref{eq:PositronMagnetic Moment}.  The most accurate electron measurement \cite{HarvardMagneticMoment2008} has a 15 times smaller uncertainty than does the most accurate positron measurement \cite{DehmeltMagneticMoment}. 
These electron and positron magnitudes agree to 1.6 standard deviations, in reasonable agreement with the CPT prediction.  The magnitudes differ by $7.2 \pm 4.3$ ppt.  However, the electron and positron moment magnitudes measured in the same trap differ by only $0.5 \pm 2.1$ ppt. \cite{DehmeltMagneticMoment}.  

A positron measurement made at the current electron precision would improve the lepton CPT test by a factor of 15. This should certainly be possible given that once in the trap, a trapped electron and positron behave in essentially the same way and can be measured to the same precision. The challenge is in loading a single positron into a trap.  We thus spent time developing a method and apparatus that could load positrons from a very safe "student" source \cite{EfficientPositronAccumulation} rather then from a source with a dangerous level of activity as was used in past experiments.   

If we achieve the goal of measuring both the electron and positron moments with an uncertainty 10 times lower than the current electron uncertainty,  then the lepton CPT test would be improved by at least a factor of 150.  Insofar as electron and positron measurements in the same apparatus may share some common systematic uncertainties that may not limit the comparison, an even bigger improvement may be possible.

\section{Most Accurate Determination of the Fine Structure Constant}
\label{sec:DeterminingAlpha}

For many years the most precise value for $\alpha$ (Eq.~\ref{eq:PredictedAlpha}) came from inserting our measured electron magnetic moment (Eq.~\ref{eq:ElectronMagneticMoment}) into  Eq.~\ref{eq:StandardModelSeries}.  Solving for $\alpha$ yields the standard model prediction for $\alpha$ in Eq.~\ref{eq:PredictedAlpha} (assuming that the standard model prediction is correct and correctly evaluated) so that   
\begin{align}
\alpha^{-1}(\mue/\mu_B) =& \, 137.035 \, 999 \, 150 \, (33). \label{eq:PredictedAlpha}
\end{align}
The fractional uncertainty in $\alpha$ is 1000 times greater than the uncertainty for the magnetic moment.

\begin{figure}[htbp!]
    \centering
    \includegraphics[width=0.45\textwidth]{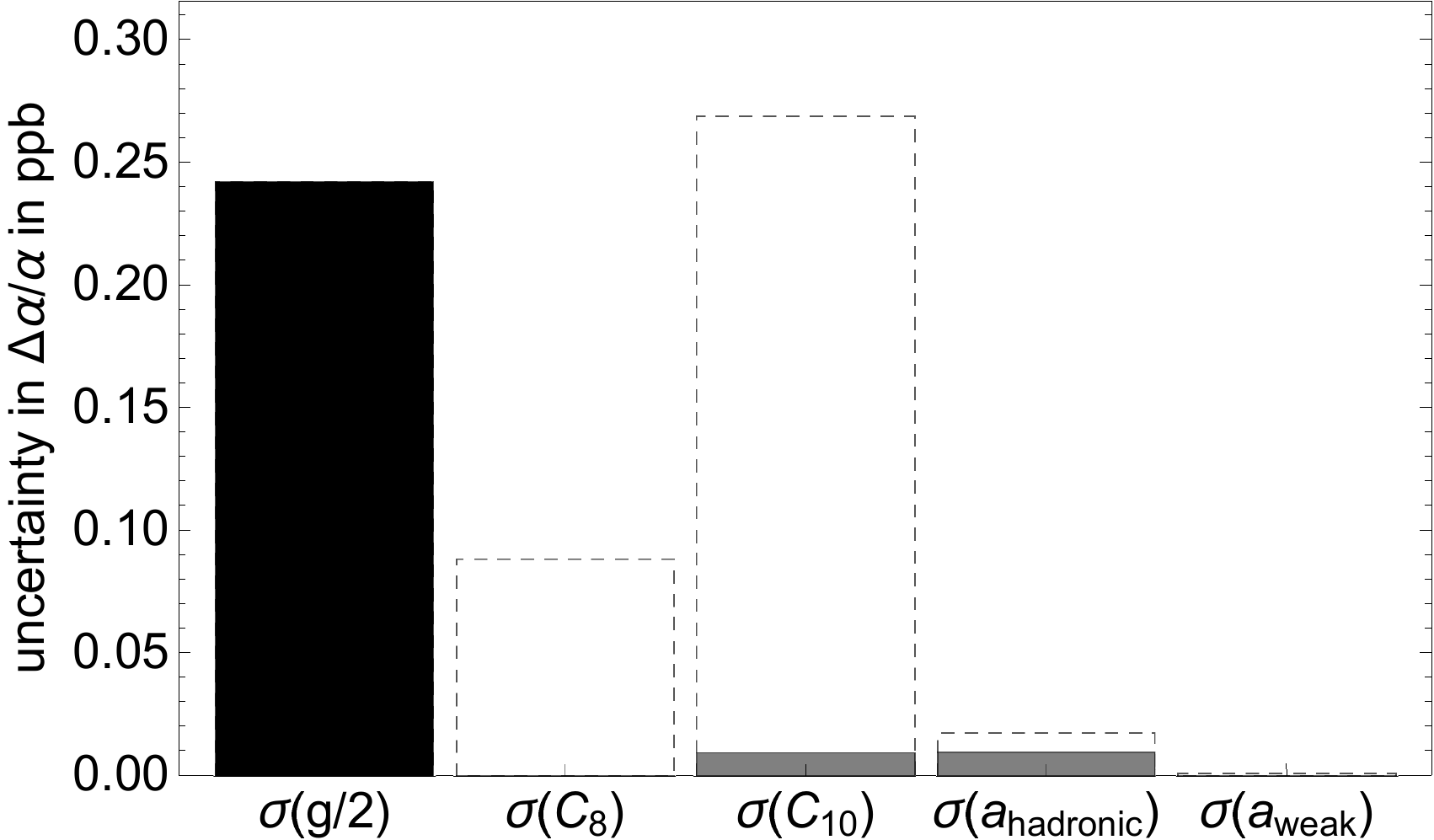}
    \caption{The uncertainty for determining $\alpha$ from a measured magnetic moment (solid) now comes mainly from the uncertainty in the moment after being limited for many years by theoretical uncertainty that was slightly larger (dashed).}
    \label{fig:AlphaUncertainty}
\end{figure}

Until recently, the uncertainty with which $\alpha$ could be determined from a measured electron magnetic moment came approximately equally from the standard model theory and our electron magnetic moment uncertainty (dashed in Fig.~\ref{fig:AlphaUncertainty}).  The recent theory advances dramatically changed the situation.  Now, the measurement uncertainty is larger than the theory uncertainty by about a factor of 15, with the theory uncertainty divided equally between the uncertainty in $C_{10}$ and $a_{hadronic}$.  A new electron measurement with a smaller uncertainty will proportionally reduce the fractional uncertainty in $\alpha$ up to about the factor of 15 without requiring better standard model theory.  A new measurement of the electron magnetic moment would again make this route the most accurate way to determine the fine structure constant.

\begin{figure}[htbp!]
    \centering
    \includegraphics[width=0.5\textwidth]{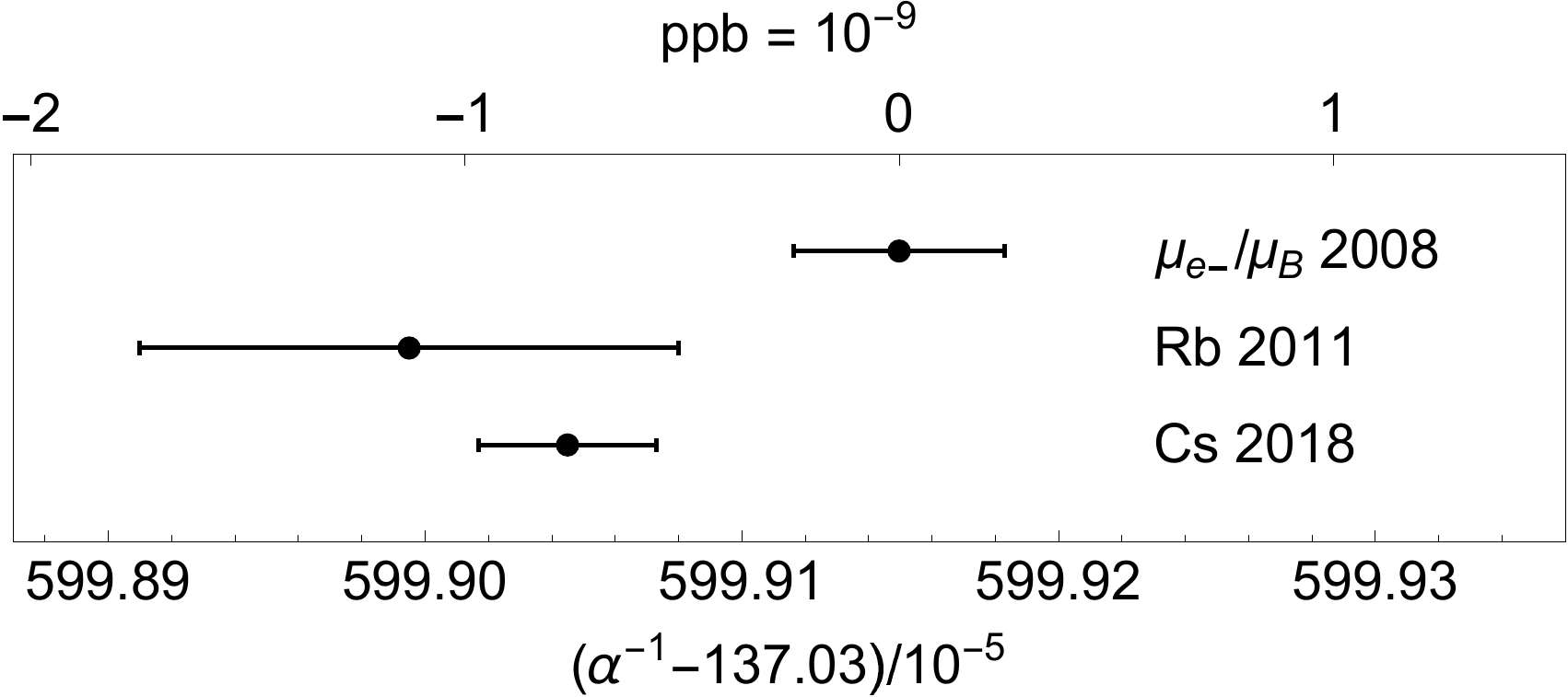}
    \caption{Most accurately determined values of the fine structure constant.  No other determinations have uncertainties small enough to fit on this graph.}
    \label{fig:AlphaValues}
\end{figure}

For testing the standard model, of course, an independent value of $\alpha$ must be inserted into   Eq.~\ref{eq:StandardModelSeries} to make the standard model prediction.  As discussed earlier, an independent measurement of $\alpha$ made just last year \ref{eq:CsAlpha} reports an accuracy comparable to ours.  
Fig.~\ref{fig:AlphaValues} compares this prediction of the standard model, using the measured magnetic moment as input, to the two other most accurately determined values of $\alpha$ in Eqs.~\ref{eq:RbAlpha} and \ref{eq:CsAlpha}. 

\section{Probe for Electron Substructure} 
\label{sec:TestingSubstructure}

Comparing experiment and theory probes for possible electron substructure at an energy scale one might only expect from a large accelerator. An electron whose constituents would have mass $m^*\gg m$ has a natural size scale, $R = \hbar/(m^*c)$. The simplest analysis of the resulting magnetic moment
\cite{BrodskyDrellElectronSubstructure} gives $\delta a \sim m/m^*$, suggesting that $m^* > 400,000~\rm{TeV}/c^2$ and $R<5 \times 10^{-25}$ m. This would be an incredible
limit, since the largest $e^+e^-$ collider (LEP) probed for a contact interaction \index{LEP contact interaction} at an $E=10.3$ TeV \cite{LepElectronStructureLimit}, with $R < (\hbar c)/E =
2 \times 10^{-20}$ m.

However, the simplest argument also implies that the first-order contribution to the electron self-energy goes as $m^*$ \cite{BrodskyDrellElectronSubstructure}. Without
heroic fine tuning (e.g., the bare mass canceling this contribution to produce the small electron mass) some internal symmetry of the electron model must
suppress both mass and moment.  For example, a chirally invariant model \cite{BrodskyDrellElectronSubstructure}, leads to $\delta a \sim (m/m^*)^2$. In this case,
$m^*>460~\rm{GeV}/c^2$ and $R<4 \times 10^{-19}$ m.  These are stringent limits to be set by an experiment carried out at $100$ mK, although they are not yet at the LEP limits. With a more precise measurement of $\alpha$, so this was limited only by our experimental uncertainty in $a$ then we could set a limit $m^*>1~\rm{TeV}/c^2$ and $R<2 \times 10^{-19}$ m.

\section{Comparison to the Muon Magnetic Moment}
\label{sec:Muon}

The electron magnetic moment is measured about $2300$ times more accurately than is the moment of its heavier muon sibling \cite{gMuon2006,HarvardMagneticMoment2008}.  Because the electron is stable there is time to isolate one electron, cool it so that it occupies a very small volume within a magnetic field, and to resolve the quantum structure in its cyclotron and spin motions.  The short-lived muon must be studied before it decays in a very small fraction of a second, during which times it orbits in a very large orbit over which the same magnetic field homogeneity realized with a nearly motionless electron cannot be maintained.

Why then measure the muon magnetic moment?  The compelling reason is that the muon
moment is expected to be more sensitive to physics beyond the standard model
by about a factor of $4\times 10^4$, which is the square of the ratio of the
muon to the electron mass.  In terms of Eq.~\ref{eq:StandardModelSeries}, such new physics would add a term $a_{\rm{new}}$ that is expected to be bigger for the muon than for the electron by this large factor, making the muon measurement a more attractive probe for such new physics.

Unfortunately, the other Standard Model contribution, $a_{\rm{hadronic}}$, is also bigger by approximately the same large factor, rather than being a much smaller correction in the electron case.  Correctly calculating these terms is a significant challenge to detecting new physics. These large terms, and the much lower measurement precision, also make the muon an unattractive candidate (compared to the electron) for determining the fine structure constant, or for testing QED..

The measured electron magnetic moment makes two contributions to using the muon system for probing for physics beyond the Standard Model. \index{Standard Model}  Both relate to determining the muon QED anomaly $a_{\rm{QED}}(\alpha)$
\begin{enumerate}
\item The electron magnetic moment measurement of is one of the two most accurate determinations of the fine structure constant. 
\item The electron magnetic moment measurement of and an independently measured value of $\alpha$ test QED calculations of the very similar terms in the electron system -- a crucial check on the calculations.
\end{enumerate}
The new physics will emerge from the muon magnetic moment measurement only if the much larger $\alpha$ dependent contributions is accurately subtracted.

\section{Other Implications of the Magnetic Moment Measurement}
\label{sec:Other}

Owing to the great accuracy of the electron magnetic moment measurements, new physics that goes beyond the standard model must always be examined to see if it would shift the electron magnetic moment.

The comparison of the measured g/2 and the value calculated from Eq. \ref{eq:StandardModelSeries} using the best available independent $\alpha$ measurement is relevant to models that attempt to explain dark matter. 
The measured g/2 is now accurate enough to essentially rule out proposed dark-matter particles with a mass that are close to the electron mass \cite{LightDarkMatterLimit}.
As another example, the ``dark photon,'' can be constrained in mass and coupling \cite{DarkPhotonFromElectrong}. The measured g/2 and a recent measurement of $\alpha$ \cite{AlphaCs2018}, dark photons with a nonzero mass are excluded up to a 99\% confidence level, restricting areas of parameter space that were not bound by accelerator experiments. The planned, increased precision, measurement allows for exploring the possibility of a dark axial vector boson, currently favored by the 2.4 $\sigma$ discrepancy with the calculated value.

A limit on a beyond-the-standard-model electron-electron coupling has been made \cite{ElectronElectronFromElectrong} using our measurement.  Our measurement also produces a bound on a beyond-the-standard-model electron-neutron coupling when combined with precision spectroscopy \cite{SpinIndependentLimits}.

\section{A New Quantum Measurement}
\label{sec:BasicIdea}

\begin{figure}[htbp!]
    \centering
    \includegraphics[width=3.5in]{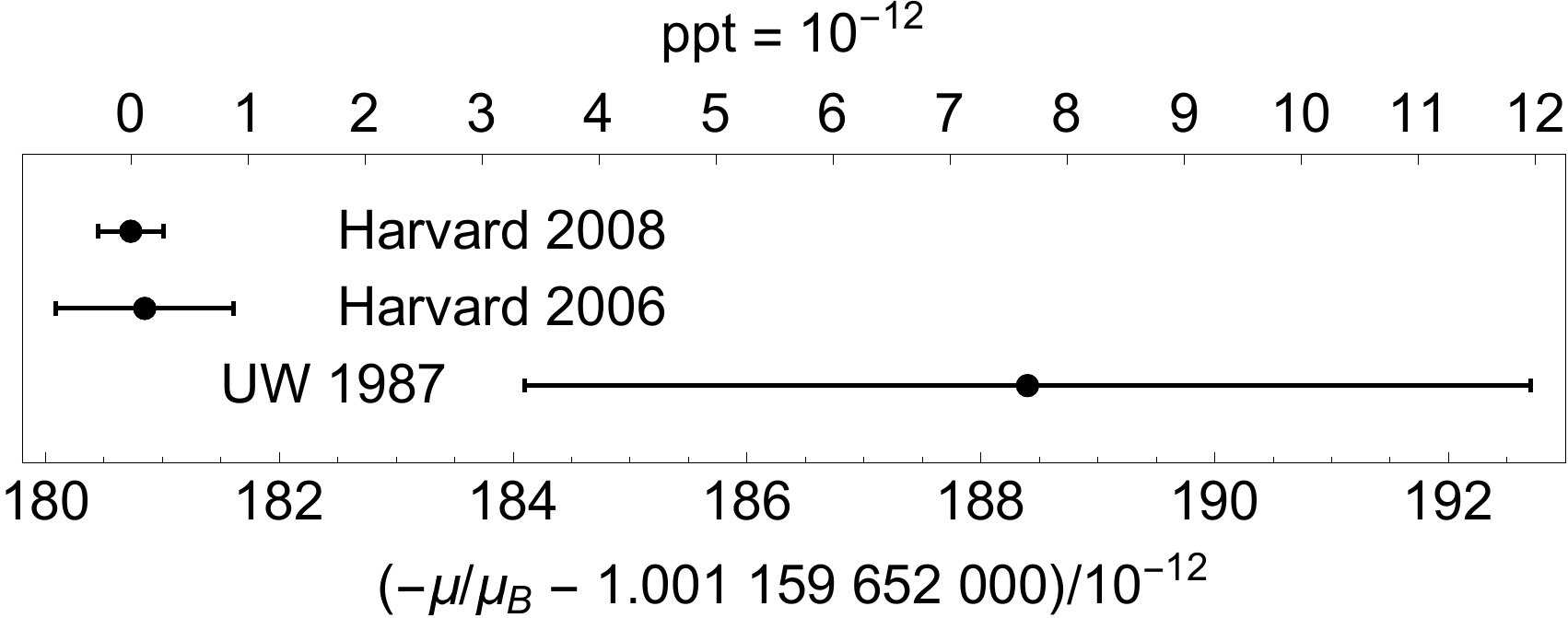}
    \caption{Comparison of the quantum determination of the electron magnetic moment with the only other measurement that fits on the same scale. }
    \label{fig:MagneticMomentMeasurements}
\end{figure}

A new quantum measurement is now beginning at Northwestern University, building on preparations at Harvard University. The goal is to improve by a factor of 10 on the earlier measurements of $\mue/\mu_B$ (Fig.~\ref{fig:MagneticMomentMeasurements}), and to measure $\mupos/\mu_B$ to the same precision - a factor of 150 smaller uncertainty than has been previously achieved as discussed earlier.  This section contrasts the completed and proposed quantum measurements of the electron and positron magnetic moments, and the aspirations for the new measurement.
Following sections provide more detail.

The completed measurement of the electron magnetic moments  \cite{HarvardMagneticMoment2008,HarvardMagneticMoment2011} is much more accurate than any other in a long history of measurements \cite{HistoryOfElectrong}.  Fig.~\ref{fig:MagneticMomentMeasurements} compares our Harvard measurements to the next most accurate measurement which has a 15 times larger uncertainty \cite{DehmeltMagneticMoment}.  The big step forward comes primarily because we carried out a quantum measurement, using quantum jump spectroscopy between clearly identified quantum states of a single electron bound to our apparatus. This required that measurements be done at 0.1 K rather than at 4.2 K or higher.  The ratio of the two frequencies obtained from the quantum jump spectroscopy yields the magnetic moment in Bohr magnetons.  the magnetic field to which both frequencies are proportional drops out; the electron thus serves as its own magnetometer.  Furthermore, the spontaneous emission of cyclotron radiation from identified quantum states is inhibited with a well understood microwave cavity formed using the electrodes of the trap that confined a single electron (discussed in Section \ref{sec:TrapCavity}).     

Common to the completed and proposed quantum measurements, the desired magnetic moment can be written as the ratio of two frequencies.  
A single electron or positron in a magnetic field, $\vec{\bf B}$, has a spin flip frequency $\nu_s$ and a cyclotron frequency, $\nu_c$.  Its magnetic moment 
\begin{equation}
\pm \frac{\mu_{e\pm}}{\mu_B} =  \frac{\nu_s}{\nu_c} = 1+\frac{\nu_s-\nu_c}{\nu_c} = 1+\frac{\nu_a}{\nu_c}.\label{eq:gFreeSpace}
\end{equation}
This is fortunate because there is nothing in physics that can be
measured more accurately than a frequency (the art of time keeping being developed being so highly developed) except for a ratio of frequencies.
Also, although both of these frequencies are proportional to the magnetic field, the field dependence drops out of the ratio. It thus needs
to be stable only on the time scale on which both frequencies can be measured, and no absolute calibration of the magnetic field is required.  The electron serves as its own magnetometer.
There is an an additional advantage in measuring the anomaly frequency $\nu_a \equiv \nu_s - \nu_c$ rather than the spin frequency,  because the spin and cyclotron frequencies differ by only a part-per-thousand, so measuring $\nu_a$ and $\nu_c$ to a precision of 1 part in $10^{10}$ gives the magnetic moment to 1 part in $10^{13}$. Of course, the frequencies in Eq.~\ref{eq:gFreeSpace} must be those measured with either the positron or electron, respectively.  

The use of quantum methods to measure both frequencies is crucial for the high accuracy.      
The energy levels for an electron in a magnetic field are
\begin{equation}
    E_{n,m_s} = m_s \, h \nu_s + (n+\frac{1}{2})\,h\nu_c
\end{equation}
with spin quantum number $m_s=\pm 1/2$ and cyclotron quantum number $n=0,1,...$ of which only the lowest two, $n=0$ and $n=1$, are used.  The frequencies are determined by driving the electron with a driving force that has the appropriate spatial symmetry, and then measuring the rate of the quantum jumps as a function of drive frequency.  Quantum methods completely eliminate any velocity dependence of the measured cyclotron frequency because only one-quantum transitions between the lowest quantum cyclotron states are measured.

A real measurement is much more intricate \cite{HarvardMagneticMoment2011}  than the basic ideas sketched here  but space does not permit a complete discussion.  In following sections we will discuss crucial apparatus and methods that lead us to believe that a new quantum measurement will yield electron and positron measurements with uncertainties lower that the current electron measurement by an order of magnitude. In the rest of this section we summarize steps being taken with a goal of a substantial reduction in uncertainties. 

Most of the uncertainty in the completed determination of the electron magnetic moment, 0.24 ppt of the 0.28 ppt, arose because the observed resonance linewidths were slightly smeared out compared to what was predicted for a perfectly stable magnetic field. The departures from the expected lineshapes were consistent with instability in the magnetic field that was too small to be observed by any other means available.  Accordingly, the new quantum measurement approach and a completely new apparatus both seek to address and mitigate magnetic field fluctuations in several very different ways.  

\begin{enumerate}
    
    \item The new approach (section \ref{sec:Simultaneous}) is to measure the spin and cyclotron transition frequencies nearly simultaneously (rather than making these measurements one after the other while the magnetic field drifts unavoidably over the weeks required to acquire the needed data). 
    
    \item A new trap and refrigerator apparatus (Section \ref{sec:Apparatus}) has the trap electrodes mechanically supported by the superconducting solenoid form so that the trap and solenoid will move together to keep the oscillation center of the trapped particle in the same magnetic field.  
    
    \item A gas $^3$He NMR probe was developed to improve the spatial stability and measure magnetic field stability (Section \ref{sec:NMR}).
    
    \item A microwave trap cavity (Section \ref{sec:TrapCavity}) is designed so that cavity sideband cooling can be used to decrease the temperature of the particle's axial motion along the trap axis, an oscillation that takes the particle through the small detection gradient deliberately introduced to allow the quantum nondemolition detection (QND) of one-quantum cyclotron and spin transitions.
    
    \item The size of the magnetic gradient (and hence the measured linewidths) are 2.3 times smaller than in the completed measurement.  
    
    \item A fallback plan is to detect the quantum transitions in a trap with the magnetic gradient after transferring them from a neighboring excitation trap with a much smaller magnetic gradient.  The challenge is to make the transfer before the cyclotron motion decays.
    
\end{enumerate}

An additional but smaller source of uncertainty in the completed measurement is from the frequency ``pulling'' of the measured particle cyclotron frequency because of the coupling of the cyclotron oscillator to the radiation modes of the trap microwave cavity.  Characterization of this coupling can be done more precisely but it was discontinued in the earlier measurement once this source of uncertainty was not limiting the uncertainty of the measurement.  The statistical uncertainty of the completed measurement was also a small contribution.  The measurements were repeated only as long as needed to make this be a small contribution.  This will naturally be done in the new quantum measurement as well.     

For these systematics-limited quantum measurements it is difficult to project what uncertainties will be obtained. Our judgement is that a factor of 10 or more should be possible with our new methods and apparatus but we do not know how to prove this ahead of making the measurement. For a new and unprecedented precision, it is always possible that an unanticipated uncertainty may show up.  In the past we proposed to measure the electron magnetic moment 10 or more times more accurately, and the new quantum methods we developed actually yielded a 15-fold improvement. We hope for a similar success after a similar investment of time and energy.   
 
So that a new quantum measurement can be carried out with a positron as well as with an electron, a positron source and a positron loading trap are included in the new apparatus. These are discussed in Section \ref{sec:PositronLoading}.

\section{Simultaneous Determination of Cyclotron and Spin Frequencies}
\label{sec:Simultaneous}

\begin{figure}[htbp!]
    \centering
    \includegraphics*[width=3in]{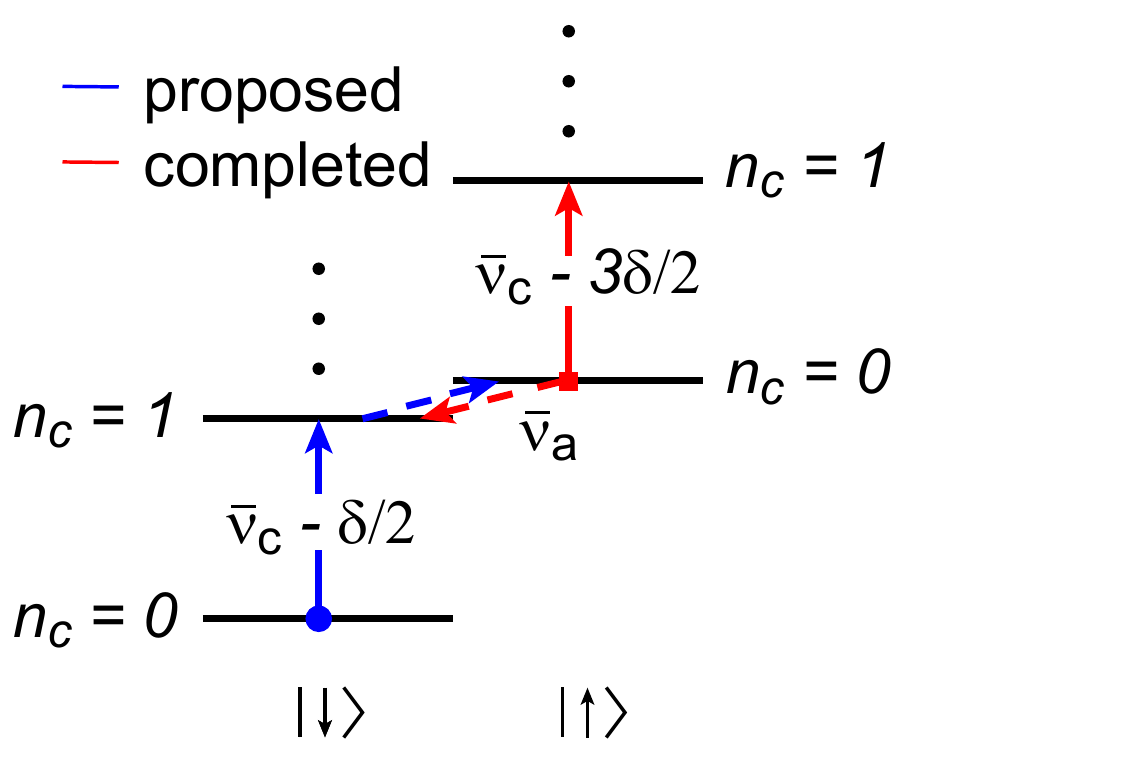}
    \caption{Lowest cyclotron, spin and axial levels of an electron in a Penning trap, including small shifts due to special relativity. Separate transitions driven for our completed measurement are in red.  For the new measurement we propose to drive the transitions in blue.   }  \label{fig:EnergyLevels}
\end{figure}

A major new idea of the new  measurement we are beginning is to drive nearly simultaneous cyclotron and anomaly transitions.  The ratio of the cyclotron and anomaly drive frequencies that is required to make a transition will be made thus nearly independent of the unavoidable drifts and fluctuations of the magnetic field. Figure~\ref{fig:EnergyLevels} contrasts how the quantum state of one particle in a Penning trap is prepared and modified for the proposed measurement, with how this was done for the completed measurement.     

For our previous measurement, we alternated between driving a cyclotron transition and then driving a spin transition. We started in the essentially stable spin up cyclotron ground state (red square).  As we drove the the cyclotron transition we watched for quantum jumps to the first excited cyclotron state (solid red arrow to the right).  We measured the spin transition frequency by driving the anomaly transition (left directed dotted red arrow) which increased the cyclotron quantum number and flipped the spin down at the same time.  Spontaneous emission then caused the system to drop to the spin down spin state.  We counted the number of spin flips from up to down as a function of the anomaly drive frequency.

For the proposed new measurement, we intend to start in the spin down cyclotron ground state (blue dot).  At the same time we propose to drive cyclotron transitions (solid blue arrow on the left) and an anomaly transition (dotted blue arrow directed right).  The feature of the proposed method is that the ratio of the two drive frequencies that causes transitions at the most rapid rate should not depend upon the unavoidable small drifts and fluctuations of the magnetic field.  The challenge of the proposed measurement is that transitions will take place when the frequency ratio is within a very narrow range of values, and at the same time that the magnetic field is at just the value that makes both of the two drives resonant with the respective transitions.

We also propose to eventually cool the axial motion of the particle because this will reduce the change in the magnetic field at the particle position that takes place as the particle oscillates through a magnetic field gradient.  The magnetic field in our trap has tiny gradients that are not perfectly shimmed out.  It also has a larger magnetic gradient deliberately introduced to make it possible to use the axial motion for quantum non-demolition detection of spin flips and one-quantum cyclotron transitions. Ideally we would like to cool the axial motion to near its quantum ground state, which will change the character of these measurements \cite{ThesisDUrso}.



\section{Trap and Refrigerator}
\label{sec:Apparatus}

\begin{figure}[htbp!]
\centering
	\begin{subfigure}{3in}
		\centering
	    \includegraphics[width=2.5in]{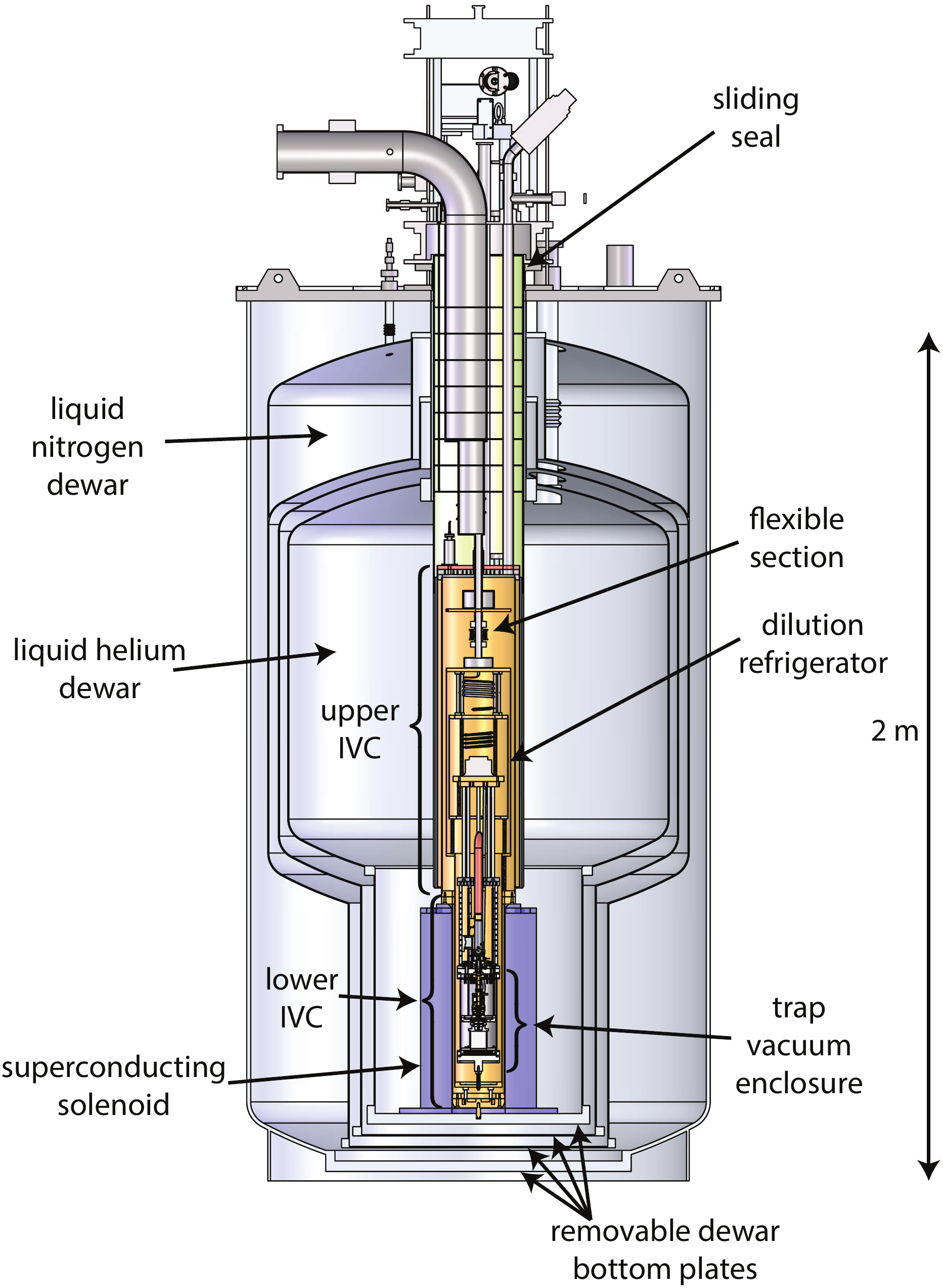}
		\caption{Overview of the  apparatus, showing the trap vacuum enclosure, solenoid, cryogen spaces and dilution refrigerator.}
		\label{fig:WholeApparatus}	
	    \end{subfigure}
    \begin{subfigure}{3in}
        \centering
    	\includegraphics[width=2.5in]{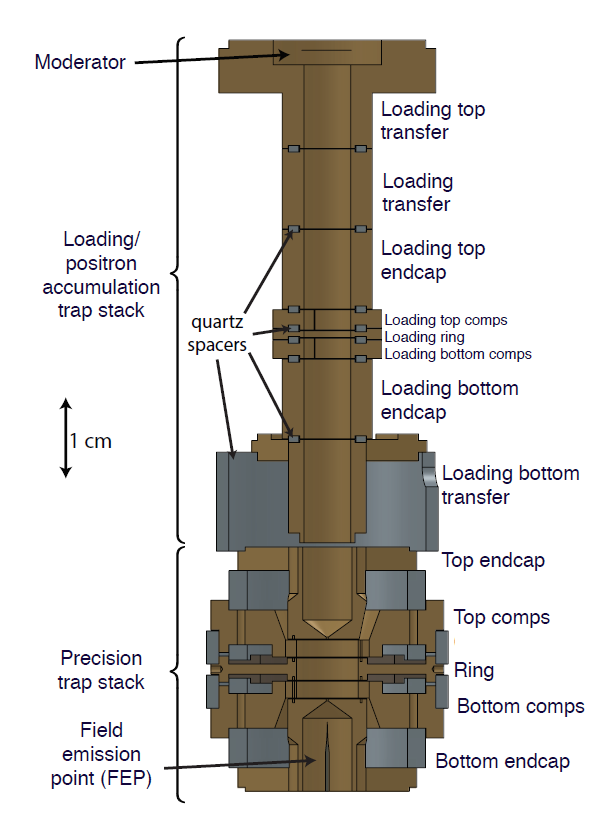}
    	\caption{Representation of the trap electrodes for positron loading and for precision measurements, each with an identifying label. }
    	\label{fig:TrapDetail}
	    \end{subfigure}
	    \caption{The cryogenic apparatus}
	    \end{figure}

The new apparatus (Fig.~\ref{fig:WholeApparatus}) is being prepared at the Center for Fundamental Physics at Northwestern University.  An initial version of the apparatus was tested first at Harvard University \cite{EfficientPositronAccumulation}. After being moved to Northwestern University, it is being improved and prepared for the new quantum measurement. It includes a 6 Tesla superconducting solenoid that can be shimmed to minimize the spatial gradients in the magnetic field.  It also includes a dilution refrigerator that cools the trap apparatus down to 0.1 Kelvin.

A crucial new apparatus feature is that the 0.1 K trap electrodes (Fig.~\ref{fig:TrapDetail}) are supported by the 4.2 K metal form upon which the superconducting solenoid is wound.  The trap hangs below the dilution refrigerator on a heat link that flexes as the trap container settles down upon the form on which the solenoid is wound. This is very different than for the apparatus used for the completed measurement, within which a trapped particle would see the magnetic field change if the trap in which it was suspended moved with respect to the solenoid producing the magnetic field. The trap and solenoid were suspended very differently for the earlier measurement. 

Two other features of this apparatus differ from that used for earlier magnetic moment measurements \cite{HarvardMagneticMoment2006,HarvardMagneticMoment2008,HarvardMagneticMoment2011}.  The center axis of the trap is aligned mechanically to the center of the superconducting solenoid by pins that center the trap within the solenoid.  To stabilize the mechanical and electrical properties of the trap we have demonstrated the regulation of the height of the liquid helium on the trap apparatus by changing the pressure in the gas above the outer part of the liquid helium dewar.

Supporting the trap from the superconducting solenoid (to keep the trap from changing its location within the superconducting solenoid) also introduces three significant cryogenic challenges.
The first cryogenic challenge is to avoid quenching the superconducting solenoid when a warm trap apparatus is initially inserted. This is crucial insofar as the field stability required for the precise measurement is only attained if the superconducting solenoid is kept at its 4.2 K operating temperature for months.  Inserting a warm trap apparatus into the solenoid must be done so as to cool the trap and dilution refrigerator to 4.2 K before they make contact with the 4.2 K solenoid.  Otherwise the solenoid will quench, and only long after the solenoid is re-energized will the needed time stability of the field be restored. 

The trap and dilution refrigerator are slowly lowered into the large helium dewar that contains the superconducting solenoid over about 4 hours.  During this time the cold helium gas that boils off from the liquid helium cools the trap from room temperature to 4.2 K.  It is critical that no air be allowed into the dewar during the insertion.  Otherwise, the paramagnetic oxygen ice that builds up within the bore of the solenoid can both keep the trap and IVC from being fully inserted and also reduce the magnetic field homogeneity.  The procedure employs the use of a cryogenic o-ring and sliding seal, surrounded by a large plastic glove-bag.  A small continuous flow of helium gas prevents air from entering the cold volume of the apparatus.

Once the vacuum enclosure for the trap rests mechanically upon the 4.2 K solenoid, it is cooled from 4.2 K to 100 mK by turning on the dilution refrigerator.  The mechanical supports are carbon fiber posts with a very low thermal conductivity.  The conduction and radiation losses of the 100 mK vacuum container for the trap are small compared to the 330 $\mu$W that the dilution refrigerator (JDR-500 from Janis Research Company) is rated to sink at 100 mK.  

The second cryogenic challenge is that liquid helium consumption has become prohibitively expensive for a dewar of this size given the significant price increases for liquid helium in recent years. When the dilution refrigerator is operating it boils off about $19\pm 1.5$ liters of liquid helium per day, along with about $24\pm3$ liters of liquid nitrogen per day.  The helium consumption drops to about $9\pm 1.5$ liters per day when the refrigerator and trap are removed to be worked on.  The expense is considerable given that the dilution refrigerator must run for many months without stopping to make the precise magnetic moment measurements.  

Replacing the nitrogen and helium reservoirs by a pulse tube refrigerator is possible in principle but this replacement could cause significant vibration that  could affect our measurements despite the solenoid and trap being mechanically connected.  Instead we installed a helium reliquefier (Cryomech PT415 with a remote motor) that turns the cold helium gas that evaporates from the liquid helium back into liquid helium. When the helium reliquefier system is running there is essentially no net liquid helium boiled off from the apparatus.  This reliquefier does include a pulse tube refrigerator but there is some mechanical isolation from the trap and it should be possible to turn it off during the most sensitive parts of precise measurements. Installed accelerometers reveal the vibration spectrum so that resonances can be identified and reduced.   

The third challenge of the cryogenic operation is preventing radiation from the warm parts of the apparatus from reaching the 100 mK trap enclosure.  The radiation must be blocked while allowing an open path to lower the radioactive source down to the 100 mK trap to load positrons, and then to retract it a distance large enough to prevent spontaneous electron loading.  To illustrate the challenge,  a 0.8 mm (1/32 inch) diameter hole that allows 300 K radiation to reach the 100 mK apparatus would provide around 200 $\upmu$W of heating -- two-thirds of the total heat load that the mixing chamber of the dilution refrigerator is specified to handle.

The solution is a series of 8 baffles and a special blocking piece that float together  on the string that supports the source capsule.  This design keeps  room temperature radiation from reaching beyond the 4.2 K stage, and the 4.2 K radiation to the 100 mK trap is very small.

\section{A Gas NMR Probe for Making a Homogeneous Magnetic Field}
\label{sec:NMR}

The magnetic field needed for the contemplated measurements must be as spatially homogeneous as is possible.  The superconducting solenoid that provides this field thus not only has a main winding, it also has 12 smaller coils labeled as z$^0$, z, z$^2$, z$^3$, x, y, zx, zy, xy, x$^2$-y$^2$, z$^2$x, and z$^2$y.  The z$^0$ coil is a smaller homogeneous magnet to sweep the center magnetic field in a fast timescale without connecting to the main coil. The others are shim coils and the labels refer to the spatial symmetry of the inhomogeneity that is primarily modified by changing the superconducting current that is persistent in this shim. 

An additional "shielding coil" is added to make the system "self-shielding" \cite{SelfShieldingSolenoid}.  When the homogeneous component of the laboratory field in which the apparatus is located changes, the winding geometry for the system was designed so that the magnetic field at the center of the system keeps this field constant.  The geometry turns flux conservation into a stable central field. 

Typically NMR probes are inserted into such solenoid systems and the currents in the persistent shim coils are adjusted to make the narrowest NMR frequency resonance.  The challenge here is that the center of this solenoid is within a liquid helium dewar.  The liquid samples typically used in NMR probes (e.g. water) will freeze, whereupon the NMR linewidth will become much to broad in frequency to be useful in shimming a magnet.  

\begin{figure}[htbp!]
\centering
	\begin{minipage}{3in}
		\centering
	    \includegraphics[width=4in]{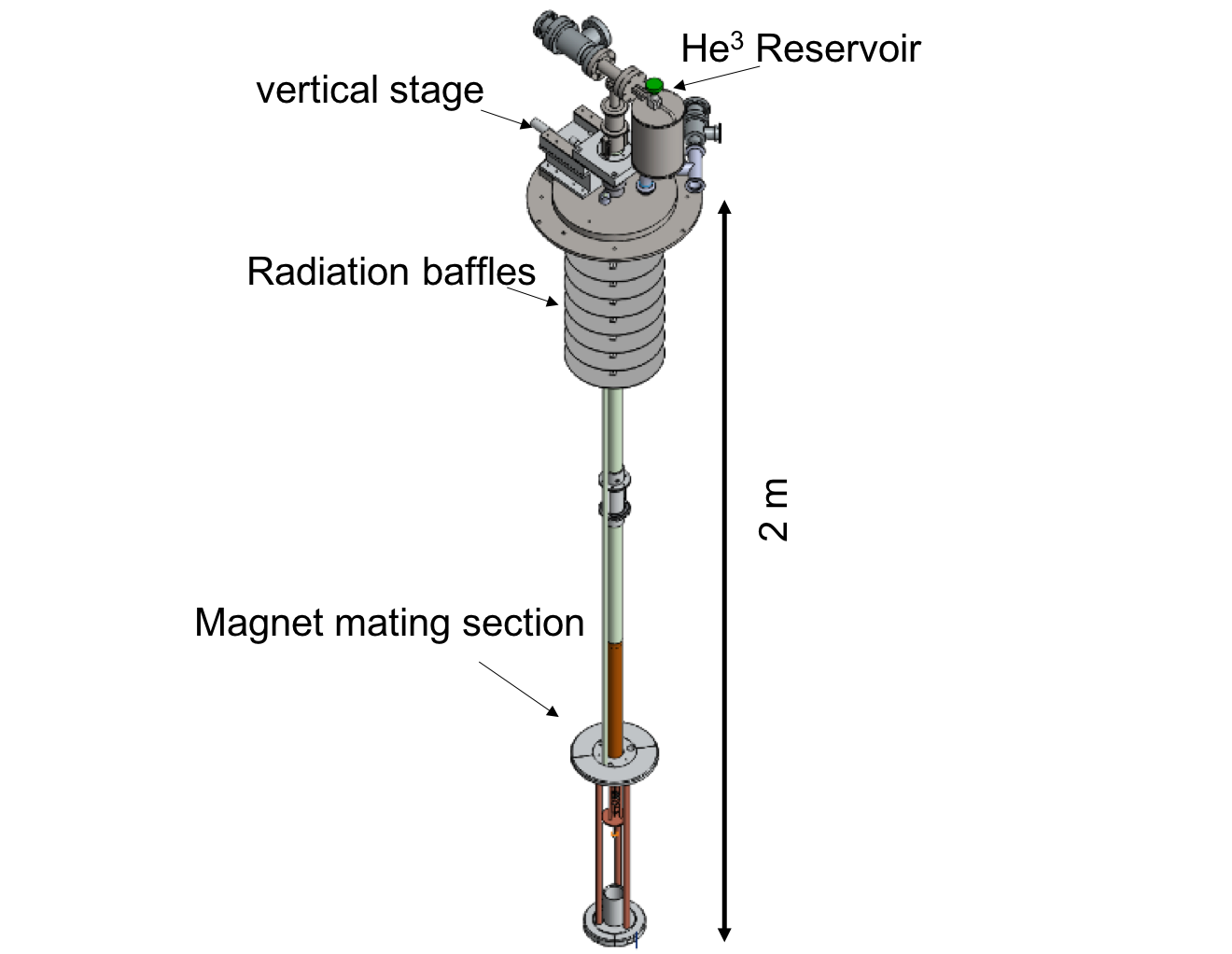}
	    \end{minipage}
    \begin{minipage}{3in}
        \centering
    	\includegraphics[width=3.5in]{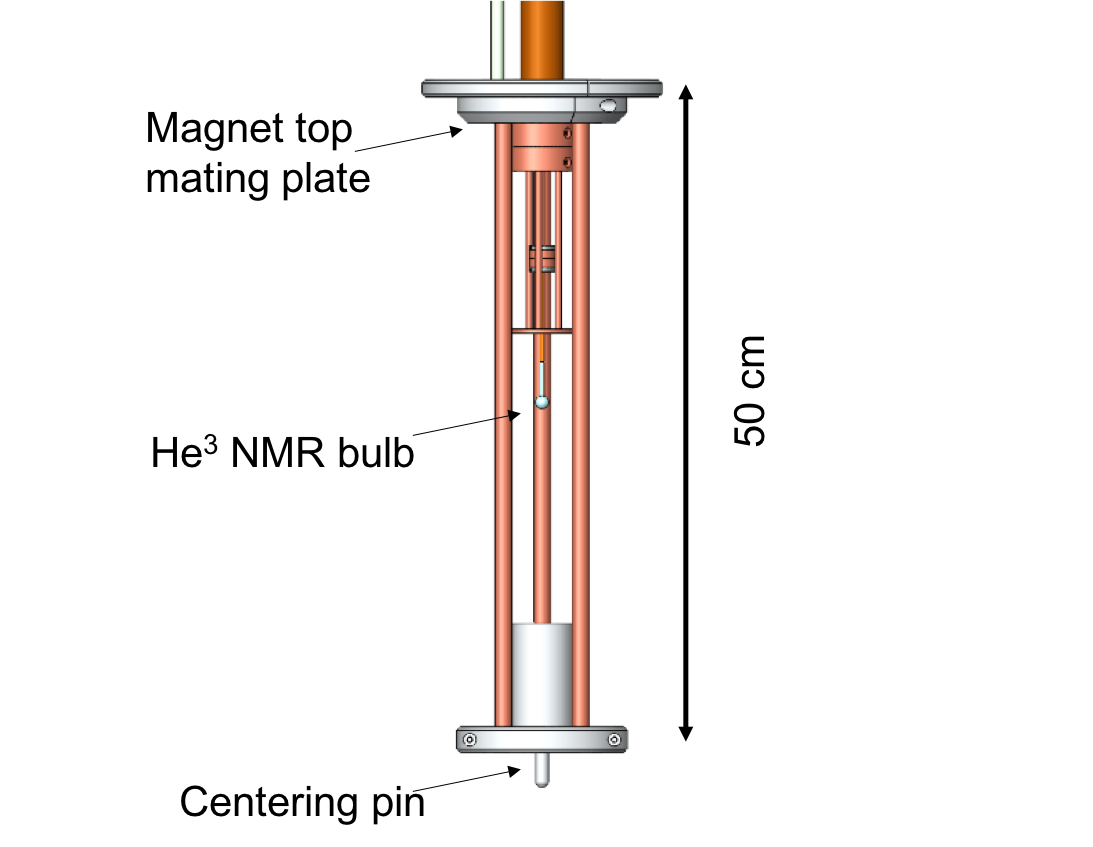}

	    \end{minipage}
	        	\caption{(left) The $^3$He NMR probe assembly is centered in the solenoid dewar via the centering pin.  The probe and be raised, lowered, and rotated.  (right) Close up of the   1 cm diameter $^3$He bulb that is connected via a capillary to a large room temperature reservoir.}
		\label{fig:NmrProbe}	
\end{figure}

Our solution is a NMR probe (Fig.~\ref{fig:NmrProbe}) that uses $^3$He gas as its active medium.  A gas typically has many fewer spins than a liquid sample  so the NMR signal size is of some concern. In order to overcome this, a large volume reservoir is prepared at room temperature so that the pressure inside the NMR bulb is kept nearly at atmospheric even at 4.2K. The radial alignment between the bulb and magnet center is given directly by the mating plates and the centering pin that tightly fit into the magnet bore. More details about this probe will be reported separately. 

Here we will focus upon the spatial homogeneity and time stability of the magnetic field of our solenoid that has been demonstrated so far using the gas NMR probe.  

\begin{figure}[htbp!]
    \centering
    \begin{minipage}{3in}
		\centering
	    \includegraphics[width=3.5in]{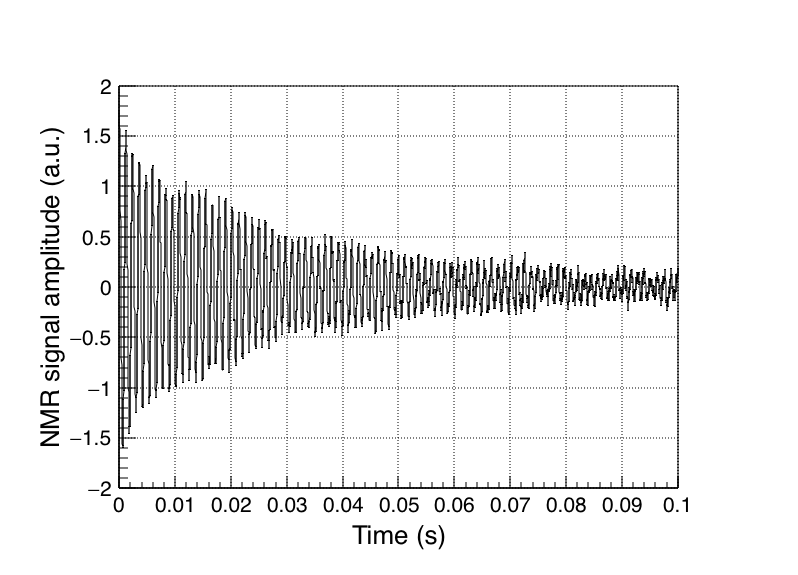}
	    \end{minipage}
    \begin{minipage}{3in}
        \centering
    	\includegraphics[width=3.5in]{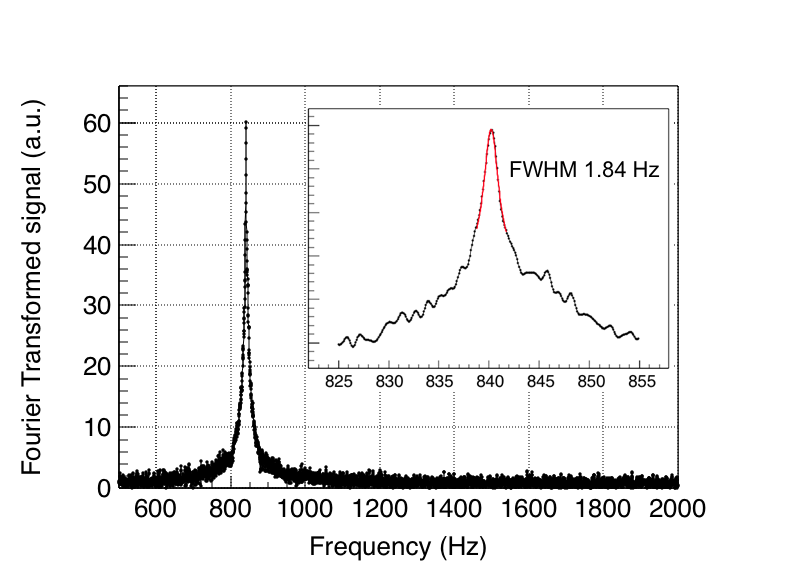}

	    \end{minipage}
    \caption{(left) The raw field induced decay signal of $^3$He at 5.3 T. The spin precession signal is amplified and mixed down before recorded. (right) Fourier transformation of the raw field induced signal. The peak at 840 Hz is the $^3$He NMR signal. Inset shows the peak and fitting by a Lorentzian function. The best fit result gives a FWHM of 1.84 Hz}.  \label{fig:NmrFIDandT2Star}
\end{figure}

After applying $\pi/2$ pulse to the $^3$He NMR bulb, the nuclear spin of each $^3$He rotates inside the vertical magnetic field. The rotation of the macroscopic magnetic moment creates a sinusoidal free induction decay signal (FID). When the magnetic field within the bulb is inhomogeneous, the $^3$He nuclei rotate at different frequencies depending on their location.  The inhomogeneity is deduced from a Fourier transform of the FID signal. For the 172.3 MHz spin precession frequency in a 5.3 T magnetic field, Fig.~\ref{fig:NmrFIDandT2Star} shows a sample FID signal mixed down to $\sim 1$ kHz (left) and its Fourier transform (right). The 1.84 Hz width of the central feature corresponds to 10.7 ppb relative inhomogeneity over 1 cm diameter spherical volume. This homogeneity is as good as the magnet used in the 2008 measurement \cite{HarvardMagneticMoment2011}.
We expect the probe to pick up some signal from helium in the capillary; this may contribute to the broad tail. These measurements are a first pass at shimming carried out only a couple of weeks after the magnetic was energized so better performance may come in the future.

\begin{figure}[htbp!]
    \centering
    \includegraphics*[width=3.5in]{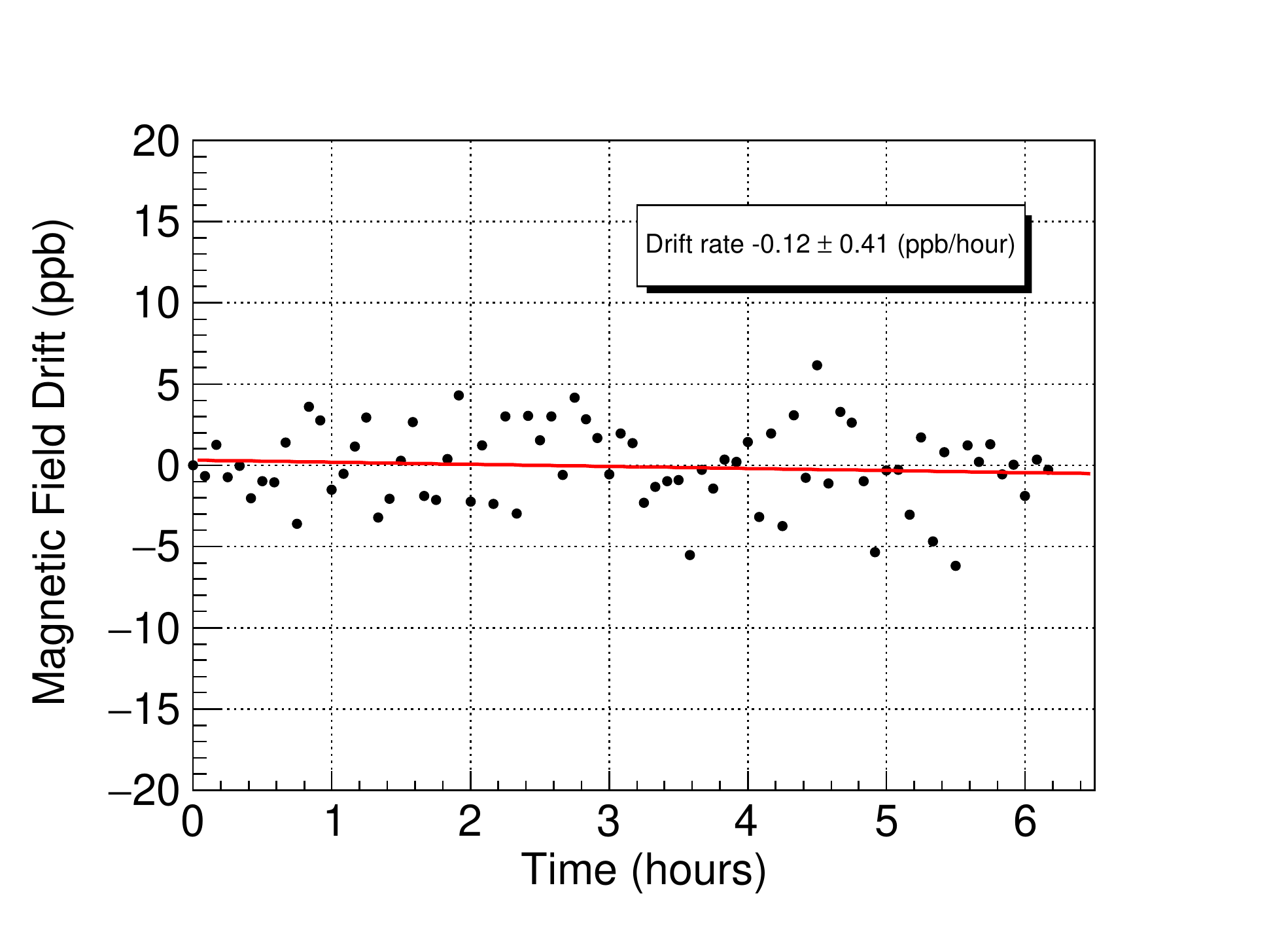}
    \caption{Measurement of the magnetic field drift with the $^3$He NMR Probe. NMR signal is taken every 5 minutes and fitted by a Lorentzian function to determine the magnetic field. The absolute magnitude of the field is 5.3 Tesla. The measured drift rate here, just 2 weeks after the solenoid was charged, is thus -0.12 $\pm$ 0.41 ppb/hour. Longer term drifts have yet to be studied.}  \label{fig:NMRDrift}
\end{figure}

The time stability of the field is critical to our proposed measurement.  To measure this, the peak of a Lorentzian fit to the top of the narrow feature is monitored as a function of time.  Only 2 weeks after the magnet was charged, the  long term drift (Fig.~\ref{fig:NMRDrift}) is already only -0.12 $\pm$ 0.41 ppb/hr.  This value is as good as the $4 \times 10^{-10}$/hr drift rate correlated with room temperature that was realized in a smaller solenoid whose dewar pressures were stabilized \cite{PbarMagneticMoment}. 
This magnet did not yet have regulated dewar pressure so there is room for improvement.  

\section{Microwave Trap Cavity for Cavity Sideband Cooling}
\label{sec:TrapCavity}

The trap electrodes are fabricated from high purity silver to minimize the effects of nuclear magnetism.  At 0.1 K, the nuclear magnetism can be surprisingly large, making it impossible to use pure copper electrodes.  The electrodes form an orthogonalized trap \cite{CylindricalPenningTrap} within which the shape of the desired electrostatic quadrupole trap potential can be optimized without large changes in the axial oscillator frequency for oscillations in the magnetic field direction. The electrodes were polished with fine polishing papers and pastes down to <0.5 $\upmu$m flatness before being gold coated.   

For the new proposed measurement we have found a cylindrical cavity geometry for which the particle cyclotron frequency can be tuned farther from resonance with cavity modes that stongly couple to cyclotron motion of a centered particle to further suppress spontaneous emission.  The same geometry produces microwave modes with the right spatial field symmetry for cavity sideband cooling while also inhibiting spontaneous emission.  The spatial field symmetry of these ``cooling modes'' should allow cavity sideband cooling of the axial motion when these modes couple axial and cyclotron motion.   

\begin{figure}[htbp!]
    \centering
    \includegraphics*[width=4.5in]{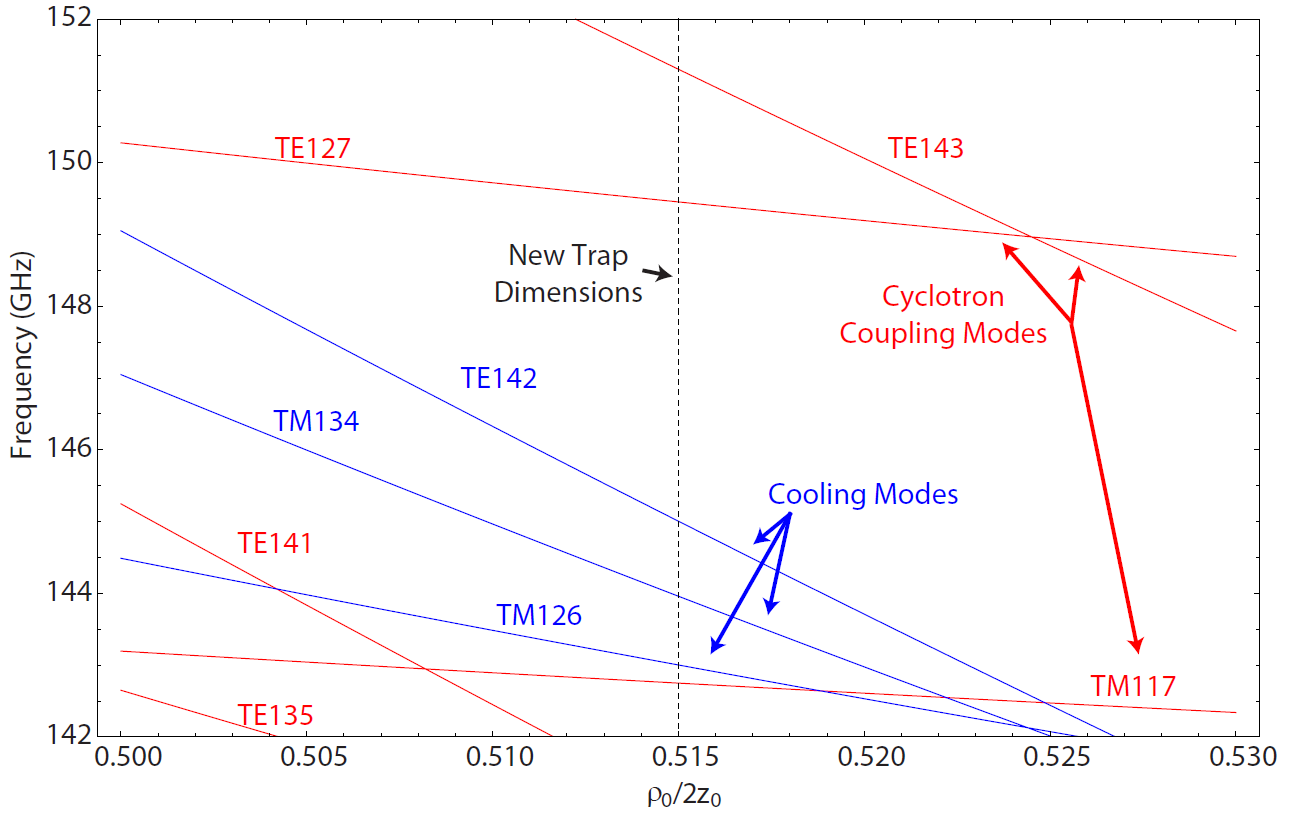}
    \caption{Calculated cavity mode structure for the current cylindrical trap cavity of radius $\rho_0$ and length $z_0$.}.  \label{fig:CavityModeStrucuture}
\end{figure}

\section{Safe and Efficient Positron Loading}
\label{sec:PositronLoading}

Once a positron is in our trap, measurements can be performed on it just as for an electron just by reversing the sign of all trap potentials.  One modern challenge in a university environment is to use a radioactive source of positrons that poses no safety risk to students.  The 0.5 mCi Na$^{22}$ source used to first trap positrons \cite{UwPositrons} was dangerous enough to require many special precautions. It is increasingly difficult  use such a source at a university given modern safety standards and expectations. By applying an unusual positron loading method we can now accumulate more than enough positrons for a precision measurement \cite{EfficientPositronAccumulation} from a source that is only $1.4$ $\upmu$Ci -- a source of the type that can be safely used by students with only modest safety precautions.  

An extremely strong source was used in the past because it was so inefficient to slow positrons that emerges from a radioactive source with an average energy of 546 keV of energy slowed to a stop within our trap. We load positrons much more efficiently by using electric field ionization of highly-excited, barely bound positronium atoms produced from a single crystal W moderator -- a method we developed for antihydrogen production \cite{PositronsFromPositronium}. These are caught within the loading trap (as shown in Fig.~\ref{fig:TrapDetail}) with a rate of 3-6 (e$^+$/s)/ $\upmu$Ci observed \cite{EfficientPositronAccumulation}. These are then quickly pulsed into the precision trap for measurement, with a small cloud used for initial diagnostics and a single positron for the final measurement. 

The Na$^{22}$ source is lowered through a tube that is bent to prevent room temperature radiation from traveling along a straight line to the 0.1 K apparatus.  The source is on a retractable line above, which allows lowering the source near to the trap only during the loading process.  In its lowered position, it emits directly into the trap can through a 10 $\upmu$m Ti window that maintains the quality of the vacuum within the trap vacuum enclosure.  The source is raised to prevent fast particles and secondary electrons from entering the trap vacuum enclosure to interfere with the precise $\mu/\mu_B$ measurement. The potential additional heat load to the dilution refrigerator is a considerable challenge, which has been mitigated by the use of a series of baffles, as mentioned above.

The positrons are captured and accumulated in a separate ``loading trap'' next to the ``precision trap'' in which the magnetic moment measurement will be made.  Loading directly into the precision trap would require an entrance aperture that is large enough to allow microwave radiation to escape from the microwave cavity formed from the cylindrical electrodes of the precision trap.  A low loss microwave cavity is required to inhibit the spontaneous emission of cyclotron radiation. Without this, we would not have the averaging time required to resolve one-quantum transitions between quantum cyclotron states.  

To minimize microwave losses from the precision trap microwave cavity, we thus introduce only two small holes in the end electrodes of the precision trap, one of which is to allow positrons accumulated in the loading trap to be transferred into the precision trap.  A fast pulsed transfer scheme is being developed to allow a high efficiency transfer through the narrow hole from the loading to precision trap.  The small hole on the other end allows electrons from a field emission point to also be loaded into the trap. To minimize microwave losses from the gaps between the trap electrodes, microwave choke groves ($\lambda$/4 at the cyclotron frequency in size) reduce the microwave power that can escape though the gaps.

\section{Conclusions}
\label{sec:Conclusion}

There are very compelling motivations for new measurement of the electron magnetic moment, and for measuring the positron moment to the same accuracy. These include testing the standard model prediction at an unprecedented precision (10 times improved) and making the most stringent test of CPT invariance with a lepton system (150 times improved), along with testing for electron substructure, setting limits on light dark matter particles and dark photons. A substantial new apparatus is being developed to make a new quantum measurement possible. Positrons can be loaded from a ``student source'' into a Penning trap with an extremely stable magnetic field, whose spatial inhomogeneity can be reduced using a gas NMR probe which works at 4 K.  The trap cavity will allow the inhibition of spontaneous emission of cyclotron radiation, and also cavity sideband cooling of axial motion.  We hope to complete new measurements within a year or two.

\acknowledgments{This work was supported by the National Science Foundation.  X. Fan is partially supported by Masason-Foundation.}

\reftitle{References}
\bibliography{main.bbl}

\end{document}